\documentclass[aps,pra,twocolumn]{revtex4}
\usepackage{graphicx}
\usepackage{amsmath,amssymb,amsthm,dsfont,bm}

\newcommand{\tr}[0]{\operatorname{tr}}  
\newcommand{\eg}[0]{\textit{e.g., }}  
\newcommand{\CP}[0]{\mathcal{CP}}  
\newcommand{\I}[0]{\mathcal{I}}  

\begin{document}
\preprint{}
\title{A resource-theoretic approach to vectorial coherence}
\author{G. M. Bosyk} 
\affiliation{Instituto de F\'{\i}sica La Plata, UNLP, CONICET, Facultad de Ciencias Exactas, 1900 La Plata, Argentina}
\author{G. Bellomo}
\affiliation{CONICET-Universidad de Buenos Aires, Instituto de Investigaci\'on en Ciencias de la Computaci\'on (ICC), Buenos
Aires, Argentina}
\author{A. Luis}
\email{alluis@fis.ucm.es}
\affiliation{Departamento de \'{O}ptica, Facultad de Ciencias F\'{\i}sicas, Universidad Complutense, 28040 Madrid, Spain}

\date{\today}

\begin{abstract}
We propose a formal resource theoretic approach to asses the coherence between partially polarized electromagnetic fields. We show that naturally defined incoherent operations endow partial coherence with a preorder relation that must be respected by all coherence measures. We examine most previously introduced coherence measures from this perspective.
\end{abstract}

\maketitle

\noindent {\em Introduction.---}
Coherence is a basic physical property that emerges in very different contexts, from classical optics to quantum mechanics. Recently, coherence has been identified as a resource for novel quantum technologies~\cite{Baumgratz2014,Cheng2015}. In classical optics coherence has two extreme physical manifestations: (i) interference, when superimposing beams with the same vibration state, and (ii) polarization, when superimposing beams with orthogonal vibration states. This makes specially attractive the analysis of coherence for the superposition of partially polarized waves where interference and polarization combine~\cite{Okoro2017}.

The complexity of the subject has motivated the introduction of several different measures of vectorial-field coherence that can actually be mutually contradicting~\cite{Karczewski1963,Gori1998,Ozaktas2002,Wolf2003,Tervo2004,Setala2004,Refregier2005,Refregier2007,Luis2007,Luis2007b,Luis2010,Luis2012,Luis2016,Abouraddy2017}. So, it emerges that a meta-theory is needed if we want to apprehend the elusive concept of coherence in this rather rich context.

We think that such comprehensive enough approach can be provided by appealing to a resource theoretic formalism, mimicking the one originally introduced for entanglement and quantum coherence~\cite{Vidal2000,Baumgratz2014}. Indeed, we have recently applied the powerful resource theory formalism to the problem of quantifying the degree of polarization of two and three dimensional random electromagnetic fields~\cite{Bosyk2017}.

We show that naturally defined incoherent operations endow partial coherence with a preorder relation that must be respected by all coherence measures. More specifically, our proposal is that  any \textit{bona fide} degree of vectorial coherence must behave monotonically with respect to the action of incoherent operations defined by the corresponding theory. We test the formalism by constructing the corresponding resource theories that arise when following the two different approaches most commonly encountered in the literature about what an incoherent and partially-polarized beam is. That is whether we are considering polarization-sensitive or polarization-insensitive coherence.

\bigskip
\bigskip

\noindent {\em Coherence-polarization state.---}
For definiteness, we focus on the vectorial electric field $\pmb{E}$ at two spatial points $\pmb{r}_1$ and $\pmb{r}_2$ with just two non vanishing components at each point, say $E_x$ and $E_y$. This can be the transverse electric fields at the pinholes of a Young interferometer. The complete system is made up of four scalar electric fields that we will consider in the space-frequency domain $E_\ell ( \pmb{r}_j , \omega )$ with $\ell = x,y$ and $j=1,2$. The dependence on the temporal frequency $\omega$ will be omitted from now on. Their second-order statistics will be completely accounted for by the cross-spectral tensor or coherence-polarization state, this is the $4 \times 4$ Hermitian nonnegative matrix $\Gamma$,
\begin{equation}
\label{eq:gamma}
 \Gamma =%
 \begin{pmatrix}
 \Gamma^{x,x}_{1,1} & \Gamma^{x,y}_{1,1} & \Gamma^{x,x}_{1,2} & \Gamma^{x,y}_{1,2} \\
 \Gamma^{y,x}_{1,1} & \Gamma^{y,y}_{1,1} & \Gamma^{y,x}_{1,2} & \Gamma^{y,y}_{1,2} \\
 \Gamma^{x,x}_{2,1} & \Gamma^{x,y}_{2,1} & \Gamma^{x,x}_{2,2} & \Gamma^{x,y}_{2,2} \\
 \Gamma^{y,x}_{2,1} & \Gamma^{y,y}_{2,1} & \Gamma^{y,x}_{2,2} & \Gamma^{y,y}_{2,2}
 \end{pmatrix}
 =
 \begin{pmatrix}
 \Gamma_{1,1} & \Gamma_{1,2} \\
 \Gamma_{2,1} & \Gamma_{2,2}
 \end{pmatrix},
\end{equation}
where the elements matrix are $\Gamma^{\ell,\ell'}_{j,j'} = \langle E_\ell (\pmb{r}_j) E_{\ell'}^*(\pmb{r}_{j'}) \rangle$ with $\ell,\ell' = x,y$ and $j,j'=1,2$, whereas the angle brackets and asterisk denote ensemble averaging and complex conjugation, respectively. Notice that in the block-matrix representation of $\Gamma$ (r.h.s of~\eqref{eq:gamma}), the matrices $\Gamma_{1,1}$ and $\Gamma_{2,2}$ represent the $2 \times 2$ Hermitian polarization coherency matrices at $\pmb{r}_1$ and $\pmb{r}_2$, respectively. On the other hand, $\Gamma_{1,2}$ and $\Gamma_{2,1}$ are the $2 \times 2$ beam coherence-polarization matrices~\cite{Gori1998}, which are non Hermitian in general but satisfy $\Gamma_{1,2} = \Gamma_{2,1}^\dagger$. The usefulness of this representation to the field statistics through $\Gamma$, instead of considering separately the submatrices $\Gamma_{j,j'}$, has been already exploited in Refs.~\cite{Luis2007,Refregier2007b,Abouraddy2014,Abouraddy2017,Gori2006}.

Let us observe that, focused to the goal of studying coherence properties, states with the same total intensity given by the $\tr \Gamma = \tr \Gamma_{1,1} + \tr \Gamma_{2,2}$ can be considered as equivalent. More precisely, let us say that $\Gamma$ is equivalent (in coherence sense) to $\Gamma'$, denoted as $\Gamma \sim \Gamma'$, if and only if $\Gamma = k \Gamma'$ for some $k>0$. Accordingly, we may assume without loss of generality that the coherence-polarization matrices of the form of~\eqref{eq:gamma} have the same trace. For simplicity we choose equal unit trace and introduce the set coherence-polarization normalized matrices $\mathcal{CP} = \left\{ \Gamma \in \mathbb{C}^{4 \times 4}: \Gamma \geq 0 \ \text{and} \ \tr \Gamma = 1 \right\}$.
\bigskip
\bigskip

\noindent {\em Coherence measures.---}
Let us distinguish between two alternative approaches. They differ on whether we include polarization in the account of coherence. Let us call them \textit{complete coherence} and \textit{interferometric coherence}. The key point is that both lead to different resources theories since they define different classes of incoherent states.

\bigskip

\noindent {\em Complete coherence.---}
Let us consider a couple of measures of total coherence. For example, we have~\cite{Luis2007}
\begin{equation}
\label{mg}
\mu_g(\Gamma) = \sqrt{\frac{4}{3} \tr\left[\left(\frac{\Gamma}{\tr \Gamma} - \frac{I_4}{4} \right)^2\right] },
\end{equation}
where $I_4$ is the $4\times 4$ identity matrix.

As another possibility motivated by optimum interferometric resolution and visibility~\cite{Luis2007b,Luis2012}, we have the measure
\begin{equation}
\label{mF}
\mu_{F,\textrm{max}}(\Gamma) = \frac{\gamma_{1} -\gamma_{4}}{\gamma_{1} + \gamma_{4}},
\end{equation}
where $ \gamma_{1}$ and $\gamma_{4}$ are the maximum and minimum of the eigenvalues of $\Gamma$, respectively.

\bigskip

\noindent {\em Interferometric coherence.---}
Regarding the interferometric-only facet of coherence, different measures of coherence have been proposed. Based on the analysis of the fringe visibility in a Young interference experiment the following quantity have been introduced as a degree of coherence~\cite{Karczewski1963,Gori1998,Wolf2003}
\begin{equation}
\label{KGW}
\mu_{KGW}(\Gamma) = \frac{\mathrm{tr} \Gamma_{1,2} }{\sqrt{\mathrm{tr} \Gamma_{1,1} \mathrm{tr} \Gamma_{2,2}}}.
\end{equation}
Accordingly, the interference fringes vanishes when $\mu_{KGW}(\Gamma)=0$, which represents the incoherence condition. However, notice that this quantity is not invariant under local unitary transformations, where by local we mean polarization.

An alternative proposal invariant under local unitary transformations has been proposed as~\cite{Tervo2004,Setala2004}
\begin{equation}
\label{TSF}
\mu_{TSF}(\Gamma) = \sqrt{\frac{\mathrm{tr}(\Gamma_{1,2}\Gamma_{1,2}^\dagger)}{\mathrm{tr}\Gamma_{1,1} \mathrm{tr}\Gamma_{2,2}}}.
\end{equation}
This quantity is not completely determined from the visibility (other measurements are necessaries).

Another approach to measure coherence properties is based on general invariance properties of $\Gamma$ under the action of local nonsingular Jones matrices~\cite{Refregier2005,Refregier2007}. Hence, the so called intrinsic degree of coherence $\mu_{S}(\Gamma)$ and $\mu_{I}(\Gamma)$ are defined as the singular values of the normalized matrix~\cite{Refregier2005}
\begin{equation}
\label{mSI}
\Gamma_{1,1}^{-1/2} \Gamma_{1,2} \Gamma_{2,2}^{-1/2}.
\end{equation}
The largest intrinsic degree of coherence, say $\mu_{S}(\Gamma)$ without loss of generality, coincides with maximal value $\mu_{KGW}(\Gamma)$ under the action of local Jones matrices~\cite{Refregier2006}.

All of these quantities are presented as suitable generalizations of visibility of interference fringes in the scalar case~\cite{Zernike1938}. This means that polarization is a kind of technical obstacle that must be avoided. To this end $\mu_{KGW}(\Gamma)$ fully disregards polarization, whereas $\mu_{TSF}(\Gamma)$, $\mu_{S}(\Gamma)$ and $\mu_{I}(\Gamma)$ sidestep it via invariance reasonings. In addition, all of them are related, since the maximum of $\mu_{KGW}(\Gamma)$ over all local unitaries can be expressed in terms of $\mu_{TSF}(\Gamma)$, $\mu_{S}(\Gamma)$ and $\mu_{I}(\Gamma)$~\cite{Gori2007,Martinez2007}.

\bigskip
\bigskip

\noindent \textit{Resource theoretic approach.---}
Our proposal is to tackle the problem of quantifying the degree of vectorial coherence by appealing to the formalism of resource theories. A formal resource theory for the vectorial coherence has to be built from the following basic components: (i) the set of incoherent states, say $\I$, (ii) a set of incoherent operations $\Lambda$, and (iii) the partially coherent states. Clearly, these three concepts are not independent to each other.
In general, one first defines the notion of being incoherent. Then, the notion of partially coherent state is defined from the negation of a incoherent one. Incoherent operations are introduced as those that leave invariant the set of incoherent states, that is, $\Lambda$ is an incoherent operation iff $\Lambda(\Gamma) \in \I$ for all $\Gamma \in \I$. At the end, one introduces the coherence monotones as functions that behave in a monotonic nonincreasing manner under the action of the incoherent operations. We postulate that any \textit{bona fide} degree of coherence has to be a coherence monotone. More precisely, let us say that $\mu$ is a degree of vectorial coherence only if $\mu: \CP \mapsto \mathbb{R}$ and $\mu(\Lambda(\Gamma)) \leq \mu(\Gamma), \ \forall \, \Gamma$, $\Lambda$. Thus, the intuition that the incoherent operations do not increase the degree of coherence is recovered. In particular, one can introduce a measure of the degree of coherence in a geometrical way as
\begin{equation}
 \mu(\Gamma) = \inf_{\Gamma' \in \I} d(\Gamma,\Gamma'),
 \label{eq:docdistance}
\end{equation}
where $d(\Gamma,\Gamma')$ is a distance or divergence that is contractive under the action of $\Lambda$ operations, that is, $d(\Lambda(\Gamma),\Lambda(\Gamma')) \leq d(\Gamma,\Gamma')$.

Finally, let us note that any coherence monotone will establish a total order among $\Gamma$. However, as this total order is not intrinsic to the structure of $\CP$, given any two partially coherent states, there may be different measures that assign contradictory values of the degree of coherence to them, that is, two measures can sort the states in a different way.

In the sequel, let us apply this formalism for the electromagnetic beam fields introducing two resource theories, one for complete coherence and the other one for interferometric coherence. In each resource theory, we will use the same symbols $\I$, $\Lambda$ and $\prec$ to identify the set of incoherent states, an incoherent operation and a hierarchy among the coherence-polarization states to be induced by the resource theory, respectively. Their meanings will be clear from the context.

\bigskip
\bigskip

\noindent \textit{Resource theory for complete coherence: majorization partial order.---}
Following~\cite{Ozaktas2002,Luis2007,Refregier2008}, an incoherent state  has to be invariant under arbitrary unitary transformations. Thus, for an incoherent state, $\Gamma$ has to be proportional to $I_4$. As a consequence, the set of all incoherent states is given by the convex set
\begin{equation}\label{eq:Isetmaj}
 \I = \left\{ \Gamma \in \CP: \Gamma = \frac{I_4}{4} \right\} .
\end{equation}
The operations $\Lambda$ that preserve $\I$ are the unital ones, which satisfy (see \eg~\cite{BengtssonBook})
\begin{equation}
\label{eq:unital}
 \Lambda\left(I_4 \right) = I_4,
\end{equation}
where $\Lambda$ should be understood as $\Lambda: \CP \mapsto \CP$.
The unital condition can be posed in an equivalent way in terms of a majorization relation between $\Gamma$ and $\Lambda(\Gamma)$ (see \eg~\cite{MarshallBook}). More precisely, one has $\Lambda(\Gamma) \prec \Gamma$ iff $\Lambda$ is unital~\cite{Chefles2002}. Here, $\Lambda(\rho) \prec \rho$ means that $\sum_{i=1}^n \lambda_i \leq \sum_{i=1}^n \gamma_i$ for $n=1,2,3$, where $\{\lambda_i\}_{i=1}^4$ and $\{\gamma_i\}_{i=1}^4$ are the eigenvalues of $\Lambda(\Gamma)$ and $\Gamma$, respectively, sorted in nondecreasing order. Moreover, according to Uhlmann's theorem~\cite{Uhlmann1970}, one has $\Lambda(\Gamma) \prec \Gamma$ iff $\Lambda(\Gamma) = \sum_k p_k U_k \Gamma U_k^\dag$, where $p_k \geq 0$, $\sum_k p_k = 1$ and $\{U_k\}$ are $4 \times 4$ unitary matrices. In other words, operations that do not increase coherence can be seen as random unitary transformations. These unitary transformations can be of the different nature as it is studied in~\cite{Abouraddy2017}. For instance, they can represent a global polarization unitary, local polarization unitaries, polarization-independent spatial unitary (\eg a beam splitter), polarization-dependent spatial unitary (\eg a polarization beam splitter) or any convex combination of them (see~\cite{Abouraddy2017} for their specific formulations).

Let us note that within this resource theory the coherence-polarization space is structured by a hierarchy given by the majorization among the states. However, the majorization relation does not provide a total order among them, because there are pairs of states, say $\Gamma, \Gamma'$, such that neither $\Gamma \prec \Gamma'$ nor $\Gamma' \prec \Gamma$ are satisfied. Majorization only provides a preorder. This means that, for every $\Gamma, \Gamma', \Gamma^{''} \in \CP$, one has: (i) $\Gamma \prec \Gamma$ (reflexivity), and (ii) if $\Gamma \prec \Gamma'$ and $\Gamma' \prec \Gamma^{''}$, then $\Gamma \prec \Gamma^{''}$ (transitivity). The antisymmetry property fails in general, but one has a weaker form, that is, if $\Gamma \prec \Gamma'$ and $\Gamma' \prec \Gamma$, then $\Gamma = U \Gamma' U^\dag$ and $\Gamma' = U^\dag \Gamma U$ with $U$ a $4 \times 4$ unitary matrix, where $U$ should be understood as $U: \CP \mapsto \CP$.

\begin{figure}[htbp]
 \centering
 \includegraphics[width=\linewidth]{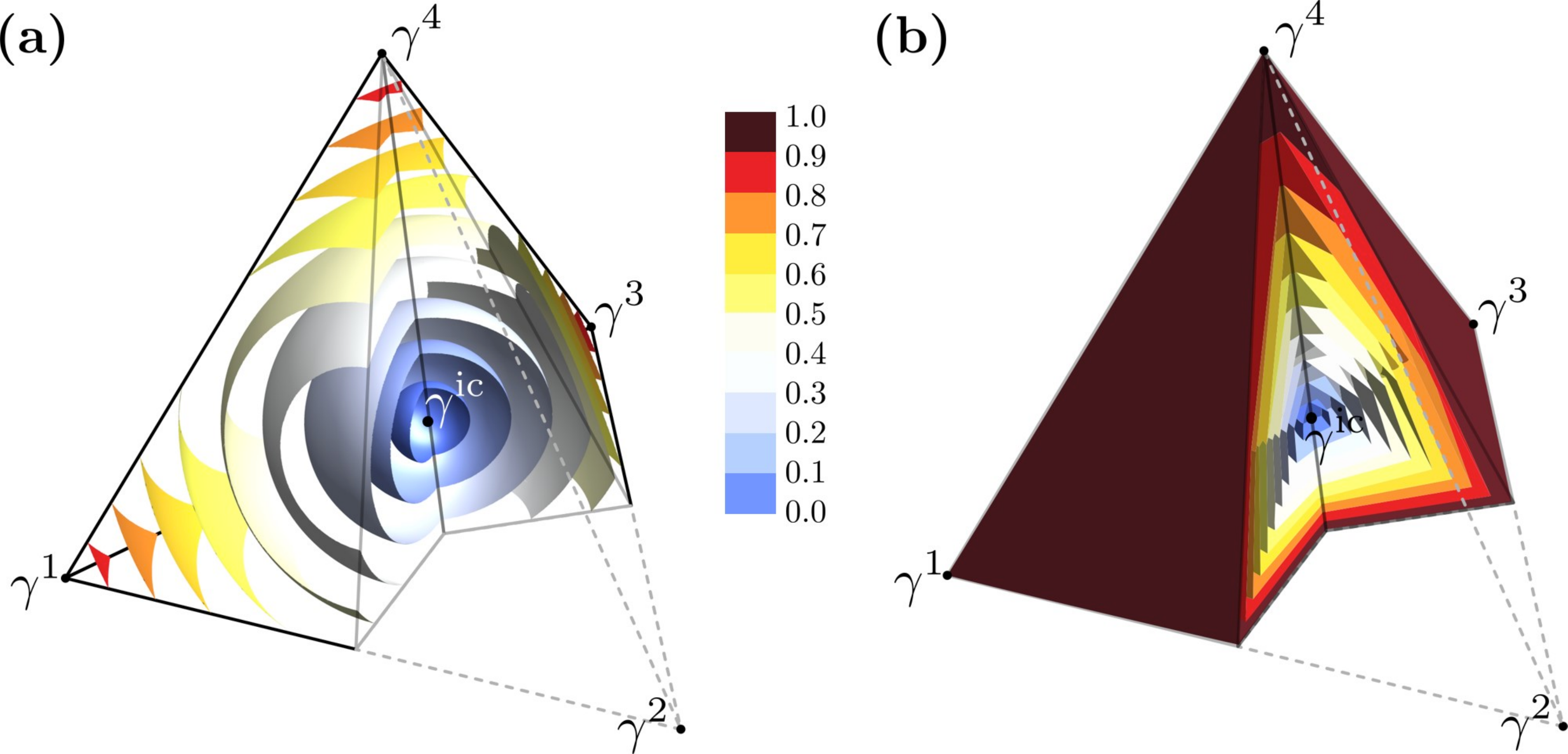}
 \caption{Isocoherence contours for (a) $\mu_g$ and (b) $\mu_{F,\max}$. The tetrahedron (or 3-simplex) gives a geometric representation of the set of probabilities vectors $\gamma = (\gamma_1,\gamma_2,\gamma_3,\gamma_4)$ given by the eigenvalues of $\Gamma$ (not necessarily sorted in a nondecreasing order). The vertices $\gamma^1=(1,0,0,0)$, $\gamma^2=(0,1,0,0)$, $\gamma^3=(0,0,1,0)$ and $\gamma^4=(0,0,0,1)$ represent maximally coherent sates, whereas the point $\gamma^{\mathrm{ic}}= \frac{1}{4}(1,1,1,1)$ represents the incoherent state. Both coherence monotones increase when going from the incoherent state to a maximally coherent one. Note that $\mu_{F,\max}$ does not distinguish $\gamma^i$ from any convex mixture of $\gamma^i$, $\gamma^j$ and $\gamma^k$, with $i,j,k=1,2,3,4$ (the faces of the the tetrahedron).}
 \label{fig:doc}
\end{figure}

The coherence monotones within this resource theory are given by Schur-convex functions, that is, functions that preserve the majorization relation: if $\Gamma' \prec \Gamma$, then $\mu(\Gamma') \leq \mu(\Gamma)$. The results in Refs.~\cite{Refregier2008,Luis2016} indicate that the measures $\mu_g(\Gamma)$ and $\mu_{F,\textrm{max}}(\Gamma)$ given by Eqs. \eqref{mg} and \eqref{mF}, respectively, are proper coherence monotones after their behavior under random unitary transformations and majorization. The behaviour of these measures can be observed in Fig.~\ref{fig:doc}. Indeed, $\mu_g(\Gamma)$ has a clear geometric interpretation as the minimum distance to the set of incoherent states. More precisely, $\mu_g(\Gamma) = \sqrt{4/3} \inf_{\Gamma' \in \I} \| \Gamma - \Gamma' \|_{HS}$, where $\| \Gamma \|_{HS} = \sqrt{\tr\Gamma \Gamma^\dag}$ stands the Hilbert-Schmidt norm of a matrix $\Gamma$.

\bigskip
\noindent \textit{Resource theory for interferometric coherence.---}
When talking about interferometric coherence, an incoherent state is one that satisfies the condition $\Gamma_{1,2} = \Gamma_{2,1} = 0_2$, with $0_2$ the $2\times2$ null matrix.  A typical physical realization holds in the case of fully random uniformly distributed relative phases between field components (see \eg~\cite{Tervo2012,Zernike1938}). Accordingly, let us introduce the set of incoherent states within this resource theory as the following convex set,
\begin{equation}
\label{eq:Isetic}
 \I = \left\{\Gamma\in\CP : \Gamma = \left(\begin{smallmatrix} \Gamma_{1,1} & 0_2 \cr 0_2 & \Gamma_{2,2} \end{smallmatrix}\right)\right\}.
\end{equation}
Here, the incoherent operations $\Lambda$ are defined as
\begin{equation}
\label{eq:icoic}
 \Lambda(\Gamma) = \frac{V\Gamma V^\dag}{\tr V\Gamma V^\dag} \ \text{with} \ V = \left(\begin{smallmatrix} V_1 & 0_2 \cr 0_2 & V_2 \end{smallmatrix}\right),
\end{equation}
where $V_{1}$ and $V_2$ are arbitrary Jones matrices. Notice that, unlike the previous resource theory, the incoherent states are not necessarily invariant under a global unitary transformation. 

The hierarchy of the different coherence-polarization states are now given by the transformations~(\ref{eq:icoic}). Let us define the binary relation: $\Gamma \prec \Gamma'$ iff there exist $\Lambda$ of the form~\eqref{eq:icoic} such that $\Gamma = \Lambda(\Gamma')$. We show that this binary relation is indeed a preorder. The reflexivity property trivially holds, because one can always choose $V_{1} = V_2 =I_2$ so that $\Gamma \prec \Gamma \ \forall \Gamma$ is satisfied. The transitivity property also holds. Notice that $\Gamma \prec \Gamma'$ and $\Gamma' \prec \Gamma{''}$ mean that there exist $\Lambda'$ and $\Lambda{''}$ incoherent operations such that $\Gamma = \Lambda'(\Gamma')$ and $\Gamma' =\Lambda{''}(\Gamma'')$. This implies that $\Gamma \prec \Gamma{''}$, because $\Gamma = \Lambda(\Gamma{''})$ with $\Lambda = \Lambda' \circ \Lambda{''}$ an incoherent operation of the form~\eqref{eq:icoic}. Again, the antisymmetric property is not satisfied in general. Instead a weaker form holds: if $\Gamma \prec \Gamma'$ and $\Gamma' \prec \Gamma$, then $\Gamma = \Lambda(\Gamma')$ and $\Gamma' = \Lambda^{-1}(\Gamma)$, where $\Lambda$ is of the form \eqref{eq:icoic} with $V_{1}$ and $V_2$ nonsingular Jones matrices.

Although $\mu_{KGW}(\Gamma)$ and $\mu_{TSF}(\Gamma)$ vanish for all incoherent states belonging to $\I$ given by Eq.~\eqref{eq:Isetic}, let us show that both measures are no proper coherence monotones of this resource theory.

First, let us consider $\mu_{KGW}(\Gamma)$. Let $\Gamma$ be given by the submatrices $\Gamma_{1,1} = \left( \begin{smallmatrix} 1/2 & 0 \cr 0 & 0 \end{smallmatrix} \right)$, $\Gamma_{2,2} = \left( \begin{smallmatrix} 0 & 0 \cr 0 & 1/2 \end{smallmatrix} \right)$ and $\Gamma_{1,2} = \left( \begin{smallmatrix} 0 & 1/2 \cr 0 & 0 \end{smallmatrix} \right)$.
Notice that $\mu_{KGW}(\Gamma)=0$, but $\Gamma \notin \I$. Even worse, let the incoherent operation $\Lambda$ defined by $V_1 = I_2$, and $V_2 = \left( \begin{smallmatrix} 0& 1 \cr 1 & 0 \end{smallmatrix} \right)$.
It can be shown that $\mu_{KGW}(\Lambda(\Gamma))=1 > \mu_{KGW}(\Gamma)=0$, so that $\mu_{KGW}(\Gamma)$ is not a coherence monotone.

Now, let us consider $\mu_{TSF}(\Gamma)$. First, let us note that $\Gamma_{1,2}\Gamma_{1,2}^\dag$ is a positive definite matrix, so that its trace vanishes iff $\Gamma_{1,2} = \Gamma_{2,1} = 0_2$. Therefore, unlike $\mu_{KGW}(\Gamma)$, we have that $\mu_{TSF}(\Gamma)= 0$ iff $\Gamma \in \I$. Now, let $\Gamma$ be defined by the submatrices $\Gamma_{1,1} = I_2/3$, $\Gamma_{2,2} = \left ( \begin{smallmatrix} 1/3 & 0 \cr 0 & 0 \end{smallmatrix} \right )$, and $\Gamma_{1,2} = \left ( \begin{smallmatrix} \mu & 0 \cr 0 & 0 \end{smallmatrix} \right )$, with $\mu > 0$. It can be shown that $\mu_{TSF}(\Gamma) = 3\mu/\sqrt2$. Let the incoherent operation $\Lambda$ be given by $V_1 = V_2 = \left ( \begin{smallmatrix} \sqrt{\lambda} & 0 \cr 0 & 1 \end{smallmatrix} \right )$, with $\lambda >0$. Then, $\mu_{TSF}(\Lambda(\Gamma)) = (3\mu \sqrt{\lambda})/\sqrt{1+\lambda}$, so that $\mu_{TSF}(\Lambda(\Gamma)) > \mu_{TSF}(\Gamma)$ if $\lambda > 1$. Therefore, we find that $\mu_{TSF}(\Gamma)$ is not a coherence monotone either.

Finally, let us examine the intrinsic degrees of coherence $\mu_S(\Gamma)$ and $\mu_I(\Gamma)$. According to Ref.~\cite{Refregier2006}, one has $\mu_S(\Gamma) = \max_{\Lambda} \mu_{KGW}(\Lambda(\Gamma))$ where $\Lambda$ are incoherent operations of the form \eqref{eq:icoic}. It is clear then that $\mu_S(\Lambda(\Gamma)) = \max_{\Lambda'} \mu_{KGW}(\Lambda'(\Lambda(\Gamma))) \leq \max_{\Lambda'} \mu_{KGW}(\Lambda'(\Gamma)) = \mu_S(\Gamma)$, given that the optimization is now performed over a restricted set of states (see Ref.~\cite{Refregier2008b} for a similar result). Moreover, one has $\mu_I(\Lambda(\Gamma))=0$ when $\Lambda$ is of the form \eqref{eq:icoic} with $V_1$ or $V_2$ singular, since $\det(\Gamma_{1,1}^{-1/2} \Gamma_{2,1} \Gamma_{2,2}^{-1/2})=\mu_S(\Gamma)\mu_I(\Gamma)$. In the case of incoherent operations with nonsingular $V_1$ and $V_2$ both intrinsic degrees of coherence remains invariant~\cite{Refregier2005}. Hence, we have proven that both quantities $\mu_S(\Gamma)$ and $\mu_I(\Gamma)$ are adequate coherence monotones for this resource theory.

As a consequence, any increasing function of the intrinsic degrees of coherence is also a coherence monotone. This holds for example when interpreting coherence as a resource for improving resolution in phase-shift detection, resolution measured for example via Cramér–Rao bound and Fisher information. For phase shifts that do not affect polarization, an interferometric coherence measure directly based on Fisher information can be introduced leading to~\cite{Luis2012}
\begin{equation}
 \mu_F(\Gamma)=\sqrt{\frac{\mu_S^2+\mu_I^2-2\mu_S^2\mu_I^2}{2-\mu_S^2-\mu_I^2}},
\end{equation}
which is, indeed, an increasing function of the intrinsic degrees of coherence. Actually, $\mu_{F,\max}$ in Eq. (\ref{mF}) can be also placed in this same metrological context as the maximum of $\mu_F(\Gamma)$ over all phase-shift schemes including those affecting polarization in a nontrivial way, in the spirit of complete coherence.

\bigskip
\bigskip

\noindent\textit{Concluding remarks.---}
In summary, we have established two resource theories for the vectorial coherence adapted to the two cases of polarization-sensitive and polarization-insensitive coherence. They define a convenient theoretical framework for the research in this subject. Furthermore, they provide a sounded criteria to validate previously introduced degrees of coherence (see Table~\ref{tab:monotones}), as well as to introduce new ones. In particular, this would rule out some of the most popular approaches considered so far.

\begin{table}[htbp]
\centering
\caption{\bf Resources theories for vectorial coherence}
\begin{tabular}{cll}
\hline
 incoherent states & incoherent operations & monotones \\
\hline
 $\frac{I_4}{4}$ & $\sum_k p_k U_k \Gamma U_k^\dag$ & $\mu_g,\mu_{F,\max}$ \\
 $\left( \begin{smallmatrix} \Gamma_{1,1} & 0_2 \cr 0_2 & \Gamma_{2,2} \end{smallmatrix} \right)$ & $\frac{V \Gamma V^\dag}{\tr V \Gamma V^\dag}$, $V = \left( \begin{smallmatrix} V_1 & 0_2 \cr 0_2 & V_2 \end{smallmatrix} \right)$ & $\mu_S,\mu_{I},\mu_F$ \\
\hline
\end{tabular}
 \label{tab:monotones}
\end{table}

\section*{Funding Information}
GMB and GB acknowledge CONICET (Argentina). AL acknowledges financial support from Spanish Ministerio de Econom\'ia y Competitividad Project No.FIS2016-75199-P, and from the Comunidad Aut\'onoma de Madrid research consortium QUITEMAD+ Grant No. S2013/ICE-2801.


\begin{thebibliography}{99}

\bibitem{Baumgratz2014}
T. Baumgratz, M. Cramer, and M. B. Plenio, ``Quantifying coherence,'' Phys. Rev. Lett. {\bf 113}, 140401 (2014).

\bibitem{Cheng2015}
S. Cheng and M. J. W. Hall, ``Complementarity relations for quantum coherence,'' Phys. Rev. A {\bf 92}, 042101 (2015).

\bibitem{Okoro2017}
C.O. Okoro, H. Kondakci, A.F. Abouraddy, and K.C. Toussaint Jr, ``Demonstration of an optical-coherence converter,'' Optica {\bf 9}, 1052--1058 (2017).

\bibitem{Karczewski1963}
B. Karczewski, ``Degree of coherence of the electromagnetic field,'' Phys. Lett. {\bf 5}, 191--192 (1963).

\bibitem{Gori1998}
F. Gori, ``Matrix treatment for partially polarized, partially coherent beams,'' Opt. Lett. \textbf{23}, 241--243 (1998).

\bibitem{Ozaktas2002}
H. M. Ozaktas, S. Y\"{u}ksel, and M. A. Kutay, ``Linear algebraic theory of partial coherence: discrete fields and
measures of partial coherence,'' J. Opt. Soc. Am. A {\bf 19}, 1563--1571 (2002).

\bibitem{Wolf2003}
E. Wolf, ``Unified theory of coherence and polarization of random electromagnetic beams,'' Phys. Lett. A {\bf 312}, 263--267 (2003).

\bibitem{Tervo2004}
J. Tervo, T. Set\"{a}l\"{a}, and A. T. Friberg, ``Degree of coherence for electromagnetic fields,'' Opt. Express {\bf 11}, 1137--1143 (2003).

\bibitem{Setala2004}
T. Set\"{a}l\"{a}, J. Tervo and A. T. Friberg, ``Complete electromagnetic coherence in the space–frequency domain,'' Opt. Lett. {\bf 29}, 328--331 (2004).

\bibitem{Refregier2005}
P. R\'{e}fr\'{e}gier and F. Goudail, ``Invariant degrees of coherence of partially polarized light,'' Opt. Express {\bf 13}, 6051--6060 (2005).

\bibitem{Refregier2007}
P. R\'{e}fr\'{e}gier and F. Goudail, ``IIntrinsic coherence: A new concept in polarization and coherence theory,'' Opt. Photon News {\bf 18}, 30--35 (2007).

\bibitem{Luis2007}
A. Luis, ``Degree of coherence for vectorial electromagnetic fields as the distance between correlation matrices,'' J. Opt. Soc. Am. A {\bf 24}, 1063--1068 (2007).

\bibitem{Luis2007b}
A. Luis, ``Maximum visibility in interferometers illuminated by vectorial waves,'' Opt. Lett {\bf 32}, 2191--2194 (2007).

\bibitem{Luis2010}
A. Luis, ``Coherence and visibility for vectorial light,'' J. Opt. Soc. Am. A {\bf 27}, 1764--1769 (2010).

\bibitem{Luis2012}
A. Luis, ``Fisher information as a generalized measure of coherence in classical and quantum optics,'' Opt. Express {\bf 20}, 24686--24698 (2012).

\bibitem{Luis2016}
A. Luis, ``Coherence for vectorial waves and majorization,'' Opt. Lett. {\bf 41}, 5190--5193 (2016).

\bibitem{Abouraddy2017} 
A. F. Abouraddy,``What is the maximum attainable visibility by a partially coherent electromagnetic field in Young double-slit interference?,''
Opt. Express {\bf 25}, 18331 (2017).

\bibitem{Vidal2000}
G. Vidal, ``Entanglement monotones,'' J. Mod. Opt. \textbf{47}, 355--376 (2000).

\bibitem{Bosyk2017}
G. M. Bosyk, G. Bellomo, A. Luis, ``Polarization monotones of 2D and 3D random EM fields'', arXiv:1709.07307 (2017).

\bibitem{Gori2006}
F. Gori, M. Santarsiero, and R. Borghi, ``Vector mode analysis of a Young interferometer,'' Opt. Lett. \textbf{31}, 858--860
(2006).

\bibitem{Refregier2007b}
P. R\'{e}fr\'{e}gier, ``Symmetries in coherence theory of partially polarized light,'' J. Phys. A {\bf 48}, 033303 (2007).

\bibitem{Abouraddy2014}
A. F. Abouraddy, K. H. Kagalwala, and B. E. A. Saleh, ``Two-point optical coherency matrix tomography,'' Opt.
Lett. \textbf{39}, 2411--2414 (2014).

\bibitem{Refregier2006}
P. R\'{e}fr\'{e}gier and A. Roueff, ``Coherence polarization filtering and relation with
intrinsic degrees of coherence,'' Opt. Lett. \textbf{31}, 1175--1178 (2006).

\bibitem{Zernike1938}
F. Zernike, ``The concept of degree of coherence and its application to optical problems,'' Physica \textbf{5}, 785--795 (1938).

\bibitem{Gori2007}
F. Gori, M. Santarsiero, and R. Borghi, ``Maximizing Young's fringe visibility through reversible optical transformations,'' Opt. Lett. \textbf{32}, 588--590 (2007).

\bibitem{Martinez2007}
R. Mart\'{\i}nez-Herrero and P. M. Mej\'{\i}as, ``Maximum visibility under unitary transformations in two-pinhole interference for electromagnetic fields,'' Opt. Lett. \textbf{32}, 1471--1473 (2007).

\bibitem{Refregier2008}
P. R\'{e}fr\'{e}gier and A. Luis, ``Irreversible effects of random unitary transformations on coherence properties of partially polarized electromagnetic fields,'' J. Opt. Soc. Am. A {\bf 25}, 2749--2757 (2008).

\bibitem{BengtssonBook}
I. Bengtsson and K. {\.Z}yczkowski, {\it Geometry of quantum states: an introduction to quantum entanglement} (Cambridge University Press, 2017).

\bibitem{MarshallBook}
A. W. Marshall, I. Olkin, and B. Arnold, {\it Inequalities: Theory of Majorization and Its Applications} (Springer, 2011).

\bibitem{Chefles2002}
A. Chefles, ``Quantum operations, state transformations and probabilities,'' Phys. Rev. A \textbf{65}, 052314 (2002).

\bibitem{Uhlmann1970}
 A. Uhlmann, ``On the Shannon entropy and related functionals on convex sets,'' Rep. Math. Phys. \textbf{1}, 147--159 (1970).

\bibitem{Tervo2012}
J. Tervo, T. Set\"{a}l\"{a}, and A. T. Friberg, ``Phase correlations and optical coherence,'' Opt. Lett. \textbf{37}, 151--154 (2012).

\bibitem{Refregier2008b}
P. R\'{e}fr\'{e}gier, ``Irreversible effects of random modulation on coherence properties of partially polarized light,'' Opt. Lett. \textbf{33}, 636--639 (2008).

\end{thebibliography}
\end{document}